\documentclass[aps,pra,reprint]{revtex4-1}
  \usepackage{newlfont}
  \usepackage{graphicx}
  \usepackage{amssymb}
  \usepackage{amsthm}
  \usepackage{amsmath,mathrsfs}
  \usepackage{bbm}
  \usepackage{braket}
  \usepackage[varg]{txfonts}
  \usepackage{xcolor}

\newtheorem*{defn}{Definition}
\newtheorem{propn}{Proposition}
\DeclareMathOperator{\Tr}{Tr}
\newcommand{\be}{\begin{equation}}
\newcommand{\ee}{\end{equation}}
\newcommand{\beq}{\begin{eqnarray}}
\newcommand{\eeq}{\end{eqnarray}}
\newcommand{\mc}[1]{\mathcal{#1}}
\newcommand{\ms}[1]{\mathscr{#1}}
\newcommand{\mb}[1]{\mathbb{#1}}
\newcommand{\mbb}[1]{\mathbbm{#1}}
\newcommand{\bv}[1]{\mathbf{#1}}
\newcommand{\bs}[1]{\boldsymbol{#1}}

\begin{document}

\title{Quantum Incompatibility of a Physical Context}

\author{E. Martins}\email{everlyn@fisica.ufpr.br}
\author{M. F. Savi}\email{msavi@fisica.ufpr.br}
\author{R. M. Angelo}\email{renato@fisica.ufpr.br}
\affiliation{Department of Physics, Federal University of Paran\'a, P.O. Box 19044,
81531-980, Curitiba, Paran\'a, Brazil}

\begin{abstract}
Pivotal within quantum physics, the concept of quantum incompatibility is generally related to algebraic aspects of the formalism, such as commutation relations and unbiasedness of bases. Recently, the concept was identified as a resource in tasks involving quantum state discrimination and quantum programmability. Here, we link quantum incompatibility with the amount of information that can be extracted from a system upon successive measurements of noncommuting observables, a scenario related to communication tasks. This approach leads us to characterize incompatibility as a resource encoded in a physical context, which involves both the quantum state and observables. Moreover, starting with a measure of context incompatibility we derive a measurement-incompatibility quantifier that is easily computable, admits a geometrical interpretation, and is maximum only if the eigenbases of the involved observables are mutually unbiased. 
\end{abstract}

\maketitle

{\it Introduction}. One of the most intriguing phenomena involving microscopic systems, quantum incompatibility is commonly associated with the noncommutativity of self-adjoint operators. This means that, contrary to the state of affairs within the classical paradigm, when two observables do not commute, their eigenvalues cannot be simultaneously obtained through a single measurement. It is then natural to take violations of joint measurability---the hypothesis that a set of measurements can be decomposed in terms of a single ``parent'' measurement---as a faithful symptom of incompatibility~\cite{Busch1986,Heinosaari2008,Ziman2016}.

Such idea has shown to be very insightful, as it unveils interconnections between the so-called measurement incompatibility and nonlocal resources, as, for instance, Bell nonlocality~\cite{Fine1982,Wolf2009,Quintino2016,Hirsch2018,Bene2018} and Einstein-Podolsky-Rosen steering~\cite{Quintino2014,Uola2014,Uola2015}. As for a quantitative assessment of the concept, incompatibility robustness measures have been introduced~\cite{Busch2013,Heinosaari2015} with a basis on the amount of noise needed to render the measurements (or devices~\cite{Haapsalo2015}) compatible. From that, further developments were accomplished within the contexts of device-independent characterizations~\cite{Cavalcanti2016,Chen2016,Chen2018}, state-discrimination tasks~\cite{Toigo2018,Toigo2019,Guhne2019,Cavalcanti2019}, and quantum programmability~\cite{Buscemi2020}, through which operational interpretations were conceived to measurement incompatibility. Recently, however, unexpected features have been noted for some widely used robustness-based measures of incompatibility~\cite{Designolle2019}.

Intuition requires that quantum incompatibility should vanish as the system approaches the classical domain---an instance that is usually accomplished through the quantum state. Accordingly, measurement incompatibility has been shown to disappear under noise~\cite{Schultz2015}. In such approach, however, one can use the duality relation $\Tr[\Gamma(\rho)\,X]=\Tr[\rho\,\Gamma_*(X)]$ to maintain a state-independent notion of measurement incompatibility. Indeed, one can always interpret any local noisy channel $\Gamma$, leading $\rho$ to a classical state, as implying some degree of fuzziness in the $X$ measurement. Nevertheless, this concept seems to be related more to experimental imperfections~\cite{Designolle2019} than to fundamental classicalization processes involving the discard of correlated systems~\cite{Zurek2007,Dieguez2018}. A subtler classical scenario can be conceived as follows. As far as heavy bodies are concerned, measurements are expected to be nearly nondisturbing, so that the resulting physical state should be independent of the ordering with which two noncommuting observables are measured. We then have a clear dependence of the notion of measurement incompatibility with an intrinsic property (the mass) of the probed body. In this case, it is less obvious how to effectively rephrase classicality in the formal structure of the measurement operators.

In this paper, by considering a scenario designed to test the safety of a communication channel, we link quantum incompatibility with information---the most fundamental resource for quantum information and quantum thermodynamics tasks~\cite{Horodecki2003,Chitambar2019,Costa2020}. Our approach employs a key principle powering quantum cryptography~\cite{Gisin2002}, namely, that no information can be extracted from a system without disturbing it~\cite{Fuchs1996}. Here, the crux is that disturbances can only occur if the measured observables and the quantum state do not commute with each other. We then introduce the concept of context incompatibility and show that it is a quantum resource for communication tasks and can be linked with a formulation of measurement-incompatibility geometry.

\begin{figure}[tb]
\includegraphics[width=\columnwidth]{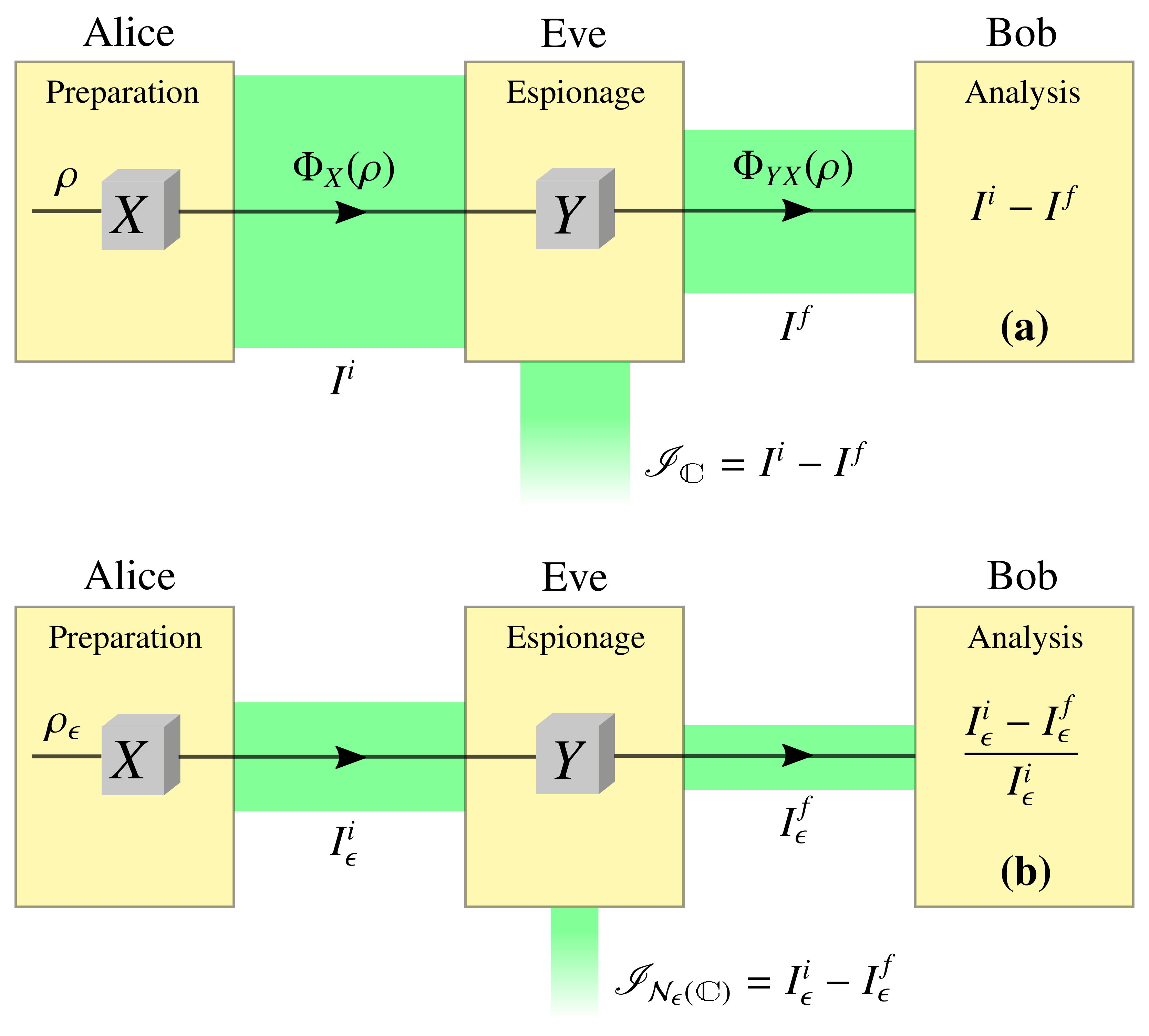}
\caption{\small (a) For a preparation $\rho$, Alice measures an observable $X$ and thus sets an amount $I^i=I(\Phi_X(\rho))$ of information (depicted by the first green thick stripe). Aware of the calibration procedure performed by Alice, the trusted partner Bob makes state tomography and then checks the information received. Upon the action of an eavesdropper, Eve, who measures $Y$, the received information actually is just $I^f=I(\Phi_{YX}(\rho))$. The incompatibility of a context $\mbb{C}\equiv\{\rho,X,Y\}$ is a resource, quantified as $\ms{I}_\mbb{C}\coloneqq I^i-I^f$, that allows for Alice and Bob to detect, via information leakage, Eve' espionage. (b) Aiming at discouraging potential eavesdroppers, Alice now prepares a highly noisy state $\rho_\epsilon=\mc{N}_\epsilon(\rho)$ [Eq.~\eqref{rhoe}] and injects a very limited amount of information $I_\epsilon^i$ in the channel. Bob receives an even more restrictive amount $I_\epsilon^f$ of information. However, because the information leakage is proportional to the injected information, by looking at the ratio $(I_\epsilon^i-I_\epsilon^f)/I_\epsilon^i$ he still succeeds to detect Eve's intervention.}
\label{fig1}
\end{figure}
%

{\it Context incompatibility}. Let $\mbb{C}\equiv\{\rho,X,Y\}\subset\mathfrak{B}(\mc{H})$ be a context such that $X=\sum_j x_j X_j$ and $Y = \sum_k y_k Y_k$ are nondegenerate discrete observables, with respective eigenbases $\text{\small $\{\ket{x_j}\}_{j=1}^{d}$}$ and $\text{\small $\{\ket{y_k}\}_{k=1}^d$}$, $X_j=\ket{x_j}\bra{x_j}$ and $Y_k=\ket{y_k}\bra{y_k}$ are projectors, $\rho$ is a generic quantum state, $\mc{H}$ is a $d$-dimensional Hilbert space, and $\mathfrak{B}\mc{(H)}$ is the set of linear bounded operators acting on $\mc{H}$. Let us consider the generic protocol depicted in Fig.~\ref{fig1}. Alice prepares a state $\rho$, with informational content
\be
I(\rho)\coloneqq\ln{d}-S(\rho),
\label{I}
\ee
where $S(\rho) = -\Tr (\rho \ln \rho)$ is the von Neumann entropy of $\rho$. After measuring $X$, without registering the outcome of any particular run of the experiment, Alice transforms the initial preparation into 
\be
\Phi_X(\rho) \coloneqq \sum_{j=1}^d X_j\,\rho\,X_j = \sum_{j=1}^d p_j X_j,
\label{mapX}
\ee
where $p_j=\Tr(\rho X_j)$. The completely positive trace preserving (CPTP) unital map $\Phi_X$ removes quantum coherence from $\rho$ in the basis $\{\ket{x_j}\}$. At this stage, the informational resource is reduced to the value $I^i\equiv I(\Phi_X(\rho))=\ln{d}-H(p_j)$, where $H(p_j)=-\sum_jp_j\ln{p_j}$ is the Shannon entropy of the probability distribution $p_j$. The system is then delivered to Bob, who expects to receive an amount $I^i$ of informational resource, as prearranged with Alice. Upon a successful verification, the trusted partners will have ascertained that the channel is safe from information leakage. Now, suppose that an eavesdropper, Eve, intercepts the system sent by Alice and probe it by measuring $Y$. The procedure is conducted by means of a unitary transformation $U\in\mathfrak{B}(\mc{H\otimes H_E})$ that entangles the system, which left Alice's laboratory in the state $\Phi_X(\rho)$, with Eve's apparatus $\mc{E}$ initially prepared in a state $\rho_0^\mc{E}\in\mathfrak{B}\mc{(H_E)}$. The composite state $\Omega_0=\Phi_X(\rho)\otimes\rho_0^\mc{E}$ thus evolves into $\Omega_t=U\Omega_0U^\dag$. Using Eq.~\eqref{I} we can  rewrite the mutual information, defined by $\mc{I}(\Omega_t)=S(\rho_t)+S(\rho_t^\mc{E})-S(\Omega_t)$, in the form $I(\Omega_t)=I(\rho_t)+I(\rho_t^\mc{E})+\mc{I}(\Omega_t)$,  where $\rho_t^\mc{E}$ is the reduced state of the apparatus and $\rho_t=\Tr_\mc{E}\Omega_t=\Phi_{YX}(\rho)$ (via the Stinespring theorem). Unitary invariance of the von Neumann entropy guarantees that $I(\Omega_0)=I(\Omega_t)$, from which we obtain $I\left(\Phi_X(\rho)\right)+I(\rho_0^\mc{E})=I\left(\Phi_{YX}(\rho)\right)+I(\rho_t^\mc{E})+\mc{I}(\Omega_t)$. With the notation $\Delta I_\mc{E}\equiv I(\rho_t^\mc{E})-I(\rho_0^\mc{E})$ and $I^f\equiv I(\Phi_{YX}(\rho))$, we arrive at
\be 
I^i-I^f=I(\Phi_X(\rho))-I(\Phi_{YX}(\rho))=\Delta I_\mc{E}+\mc{I}(\Omega_t),
\label{inc-DE}
\ee
where $\Phi_{YX}(\rho)=\sum_{j,k}\wp_{k|j}p_{j}\,Y_k$, $\wp_{k|j}=|\braket{x_j|y_k}|^2=\Tr(X_jY_k)$, and $I^f=\ln{d}-H\big(\sum_j\wp_{k|j}p_j\big)$. Clearly, the resource consumed from Alice's system, $\ms{I}_\mbb{C}\equiv I^i-I^f$, was used to change the local information of Eve's apparatus and to increase the correlations between the system and the apparatus. If $\ms{I}_\mbb{C}>0$, Alice and Bob then discover that the channel is being spied upon [Fig.~\ref{fig1}(a)]. Now, using $\sum_kY_k^2=\mbb{1}$ and $Y_k\Phi_Y(\sigma)=\Phi_Y(\sigma)Y_k$, one shows that $\Tr[\sigma g\big(\Phi_Y(\sigma)\big)]=\Tr[\Phi_Y(\sigma) g\big(\Phi_Y(\sigma)\big)]$ for any state $\sigma$ and  function $g$. This allows us to write $S(\sigma||\Phi_Y(\sigma))=S(\Phi_Y(\sigma))-S(\sigma)$, with $S(\sigma||\varrho)=\Tr[\sigma(\ln\sigma-\ln\varrho)]\geq 0$ being the relative entropy (equality holding if and only if $\sigma = \varrho$). We then arrive at the form
\be 
\ms{I}_\mbb{C}=I(\Phi_X(\rho))-I(\Phi_{YX}(\rho))=S(\Phi_X(\rho)||\Phi_{YX}(\rho)),
\label{I_C}
\ee 
through which we can check that there are only two instances in which $\ms{I}_\mbb{C}=0$: (i) $[X,Y]=0$ ($\forall \rho$) and (ii) $\Phi_X(\rho)=\mbb{1}/d$ $(\forall Y)$. In case (i), the operators share the same set of eigenstates and $\Phi_{YX}(\rho)=\Phi_X(\rho)$. Case (ii) implies that $I^i=I^f=0$. On the other hand, the consumed resource $\ms{I}_\mbb{C}$ reaches its maximum value, $\ln{d}$, when $\rho=X_j$ (an eigenstate of $X$) and, in addition, the $X$ and $Y$ eigenbases form mutually unbiased bases (MUB)~\cite{Durt2010}, that is, $|\braket{x_j|y_k}|^2=1/d$. Therefore, from Bob's (Eve's) viewpoint, noncommutativity and $I^i>0$ are necessary ingredients---resources---for a successful leakage detection (information acquisition). Thus, with respect to the protocol depicted by Fig.~\ref{fig1}(a), the following concept is introduced.

\begin{defn}
Context incompatibility is the resource encoded in a context $\mbb{C}\equiv\{\rho,X,Y\}$ that allows one to test the safety of a communication channel against information leakage. Quantified via $\ms{I}_\mbb{C}=I^i-I^f$ [Eq.~\eqref{I_C}], it is operationally related to the amount of information subtracted from the system upon an external measurement.
\label{def1}
\end{defn}

Before proceeding with the proof that context incompatibility can be framed in the formal structure of a resource theory, it is interesting to note that a connection can be made with quantum coherence---a well-established quantum resource quantified by the $\{\ket{y_k}\}$-basis relative entropy of coherence, $C_Y(\rho)=S(\rho||\Phi_Y(\rho))$~\cite{Baumgratz2014,Streltsov2017}. As one can readily check, $\ms{I}_{\mbb{C}}=C_Y(\Phi_X(\rho))$, meaning that context incompatibility can be viewed as the amount of $Y$ coherence that is encoded in an $X$-incoherent state $\Phi_X(\rho)$. This is how the incompatibility of the set $\{X,Y\}$ is captured by $\ms{I}_\mbb{C}$. 

{\it Resource theory of context incompatibility}. We now formally characterize context incompatibility as a resource. Unlike the usual account of resource theories, where the resource is encoded in the quantum state, here the resource is encoded in the whole physical context $\mbb{C}$. Following standard approaches~\cite{Chitambar2019,Costa2020}, we devise a formal structure composed of  (i) resourceless contexts, (ii) resourceful contexts, (iii) free operations, and (iv) a resource monotone. The last object is naturally identified with the measure \eqref{I_C}, but any other contractive distance function involving the states $\Phi_{YX}(\rho)$ and $\Phi_X(\rho)$ would work as a proper monotone. The resourceless (free) contexts, defined as $\mbb{C}^\text{free}$ such that $\ms{I}_{\mbb{C}^\text{free}}=0$, are
\begin{subequations}
\begin{align}
&\mbb{C}_1^\text{free}=\{\rho,X,Y\}\,\,\text{s.t.}\,[X,Y]=0\,\,(\forall\,\rho),\\
&\mbb{C}_2^\text{free}=\{\rho,X,Y\}\,\,\text{s.t.}\,\Phi_X(\rho)=\mbb{1}/d\,\,(\forall Y).
\end{align}\label{Cfree}
\end{subequations}
Notice that for $\mbb{C}_{2}^\text{free}$ one has $I^i=0$. The proof that the above are the only existing free contexts is given in the Supplemental Material~\cite{SuppMat}. Apart from them, any other context is termed resourceful. With regard to free operations, since the relative entropy is nonincreasing under generic CPTP maps $\Gamma$, it follows that $\ms{I}_\mbb{C}\geq \ms{I}_{\Gamma(\mbb{C})}$, where $\Gamma(\mbb{C})\equiv\{\Gamma(\rho),X,Y\}$, provided that $\Gamma$ commutes with the maps $\Phi_X$ and $\Phi_{YX}$. In this case, it is clear that resource is never created upon the action of $\Gamma$. Also, to ensure that $\ms{I}_{\Gamma(\mbb{C}^\text{free})}=0$, we need to require $\Gamma$ to be unital, so as not to make $\mbb{C}_1^\text{free}$ resourceful upon $\Gamma$. Altogether, these aspects characterize the free operations $\Gamma$ with respect to context incompatibility. In our approach we do not admit any operations on $\{X,Y\}$, as this would imply aspects of measurement fuzziness that have been disregarded from the outset. 

{\it Measurement incompatibility}. Let us come back to the protocol, now considering a noisy scenario [Fig.~\ref{fig1}(b)]. To discourage any potential eavesdroppers, Alice introduces, in a controllable way, an amount $1-\epsilon$ of noise in the input state, which then reads
\be 
\rho_\epsilon=\mc{N}_\epsilon(\rho)\coloneqq (1-\epsilon)\text{\small $\frac{\mbb{1}}{d}$}+\epsilon\,\rho,
\label{rhoe}
\ee 
where $\epsilon\in[0,1]$ and $\mc{N}_\epsilon$ is a CPTP unital noise map. From the concavity of the entropy and the joint convexity of the relative entropy, one can check that $I^i_\epsilon\equiv I(\Phi_X(\rho_\epsilon))\leq \epsilon\,I^i$ and $\ms{I}_{\mc{N}_\epsilon(\mbb{C})}\leq \epsilon\,\ms{I}_{\mbb{C}}$, where $\mc{N}_\epsilon(\mbb{C})=\{\rho_\epsilon,X,Y\}$ and $\mc{N}_{\epsilon=0}(\mbb{C})=\mbb{C}$. Hence, the preparation $\rho_\epsilon$ implies, for $\epsilon\ll 1$, a very limited amount of information in the channel and an equally restrictive amount of consumable information. Aware of the amount of noise introduced, Bob can still check the security of the channel by looking at the amount of information that leaks per unit of injected information. Bob computes the ratio $\ms{R}_{\mc{N}_\epsilon(\mbb{C})}\coloneqq (I_\epsilon^i-I_\epsilon^f)/I_\epsilon^i=\ms{I}_{\mc{N}_\epsilon(\mbb{C})}/I_\epsilon^i$, with $I_\epsilon^f\equiv I(\Phi_{YX}(\rho_\epsilon))$, since, in the large-noise limit, it reads
\be
\lim_{\epsilon\to 0}\ms{R}_{\mc{N}_\epsilon(\mbb{C})}=\frac{|| \Phi_{YX}(\rho)-\Phi_X(\rho) ||^2}{||\Phi_X(\rho)-\mbb{1}/d||^2}=:\ms{R}_\mbb{C},
\label{R}
\ee 
where $|| A ||\coloneqq\sqrt{\Tr(A^{\dag}A)}$ is the Hilbert-Schmidt norm of $A$. The limit is calculated as follows. Since $\ket{x_j}$ are eigenstates of $\Phi_X(\rho_\epsilon)$ with eigenvalues $\frac{1-\epsilon}{d}+\epsilon p_j$, we can explicitly compute the von Neumann entropy and the associated information, $I_\epsilon^i$. For $d\ll 1/\epsilon$, we expand the formulas and retain only terms up to order $\epsilon^2$. With this procedure, we find $I_\epsilon^i\cong\tfrac{\epsilon^2}{2}(d||\Phi_X(\rho)||^2-1)$ and $I_\epsilon^f\cong \tfrac{\epsilon^2}{2}(d||\Phi_{YX}(\rho)||^2-1)$. From the result $\Tr[\Phi_X(\rho)\Phi_{YX}(\rho)]=||\Phi_{YX}(\rho)||^2$, one is able to show that $||\Phi_{YX}(\rho)-\Phi_X(\rho)||^2=||\Phi_X(\rho)||^2-||\Phi_{YX}(\rho)||^2$. The emerging expression for $\ms{I}_{\mc{N}_\epsilon(\mbb{C})}/I_\epsilon^i$ results to be $\epsilon$-independent and the limit trivially follows. Therefore, by use of this ratio, Bob can still check information leakage for arbitrarily large noise. An interesting feature of the formula \eqref{R} is that it is invariant upon noise maps of the form~\eqref{rhoe}, that is, $\ms{R}_{\mc{N}_\epsilon(\mbb{C})}=\ms{R}_\mbb{C}$, for all $\rho$ and $\epsilon\in[0,1]$. This allows us to write, up to order $\epsilon^2$,
\be 
\ms{I}_{\mc{N}_\epsilon(\mbb{C})}\cong \ms{R}_\mbb{C}I_\epsilon^i,
\ee 
which shows that the amount of information that is extracted by Eve is directly proportional to the injected information [as suggested by the green thick stripes in Fig.~\ref{fig1}(b)]. Being $\epsilon$-independent, the proportionality ratio $\ms{R}_\mbb{C}$ might be expected to be more closely associated with the algebraic relations between $X$ and $Y$ solely (this will be shown to be true for any qubit context), but in general it provides an estimate for the context incompatibility, though in a norm-based way.

In search of a link between context incompatibility and measurement incompatibility, the natural move is to restrict ourselves to the context $\mbb{C}_j\equiv\{X_j,X,Y\}$. Then, setting $\rho=X_j$ in Eq.~\eqref{rhoe}, we find $\ms{R}_{\mbb{C}_j}=\frac{d}{d-1}\left(1-||\Phi_Y(X_j)||^2 \right)$, which is just the linear entropy of $\Phi_Y(X_j)=\sum_k\wp_{k|j} Y_k=\sum_k\Tr(Y_kX_j)Y_k$, and $I_\epsilon^i\cong \epsilon^2(d-1)/2$. It follows that $\ms{I}_{\mc{N}_\epsilon(\mbb{C}_j)}\cong \ms{R}_{\mbb{C}_j}I_\epsilon^i$, with $I_\epsilon^i$ keeping no dependence on the input state $X_j$. This result is relevant because it shows that the ratio $\ms{R}_{\mbb{C}_j}$, which is an easily computable measure, suffices to capture the level of incompatibility in the context $\mbb{C}_j$. However, it cannot be our definitive figure of merit for quantifying the measurement incompatibility of the set $\{X,Y\}$, since it considers only a single element of the $X$ eigenbasis. We then examine the averaging
\be 
\ms{M}_{\{X,Y\}}\coloneqq \text{\small $\frac{1}{d}$}\sum_{j=1}^d\ms{R}_{\mbb{C}_j}.
\ee 
By construction, $\ms{M}_{\{X,Y\}}$ tends to be a more appropriate quantifier of measurement incompatibility, for it (i) encompasses the contribution of all $X$ eigenstates and (ii) is symmetrical upon the ordering permutation $X\leftrightarrow Y$, that is, $\ms{M}_{\{X,Y\}}=\ms{M}_{\{Y,X\}}$ (a desirable property for a measure meant to describe an algebraic relation between two observables). This point can be checked from the manipulated form
\be 
\ms{M}_{\{X,Y\}}=\text{\small $\sum_{j,k=1}^d$}\frac{||\,[X_j,Y_k]\,||^2}{2(d-1)}=\text{\small $\frac{1}{d-1}$}\left(d-\text{\small $\sum_{j,k=1}^d$}|\braket{x_j|y_k}|^4\right),
\ee 
The first equality makes it explicit a relation with the commutator $[X,Y]=\sum_{j,k}x_jy_k[X_j,Y_k]$. This is an important reference to the well-known fact that, when projective measurements are concerned, joint measurability and commutativity turn out to be equivalent notions, although this is not true in general~\cite{Ziman2016}. Moreover, we have $0\leq \ms{M}_{\{X,Y\}}\leq 1$, with the upper (lower) bound being reached for, and only for, MUB (commuting operators). As demonstrated in the Supplemental Material~\cite{SuppMat}, $\ms{M}_{\{X,Y\}}$ can be derived via an independent purely algebraic  construction, which further reinforces the claim that this can be taken as a reasonable quantifier of measurement incompatibility. 

{\it Geometrical interpretation}. We now build a geometrical picture for the incompatibility measures introduced above. To this end, we employ the generalized Bloch representation, which is based on the observation that the set of matrices $\{\mbb{1},\Lambda_1,\cdots,\Lambda_{d^2-1}\}$ form a basis for linear operators acting on the state space, where the $d\times d$ complex, traceless, orthogonal, self-adjoint matrices $\Lambda_i$ are the generators of $\mathrm{SU}(d)$, the special group of degree $d$ (see the Supplemental Material~\cite{SuppMat} for a very brief review of the complete formalism~\cite{Aerts2014,Aerts2016}). With the normalization $\Tr(\Lambda_i\Lambda_j)=2 \delta_{ij}$, one can always express a quantum state as
\be 
\rho_\bv{r}=\tfrac{1}{d}\big(\mbb{1}+C_d\bv{r}\cdot\bs{\Lambda}\big),
\label{rho_r}
\ee 
where $\bv{r}=\sum_{i=1}^{d^2-1}r_i\bv{e}_i$, $\bs{\Lambda}=\sum_{i=1}^{d^2-1}\Lambda_i\bv{e}_i$, $\{\bv{e}_i\}_{i=1}^{d^2-1}$ is an orthonormal basis in $\mb{R}^{d^2-1}$, and $C_d=\sqrt{d(d-1)/2}$. Through the above parametrization, any quantum state is represented by the vector $\bv{r}$ in a $(d^2-1)$-dimensional real ball $B(\mb{R}^{d^2-1})$. Projection operators admit the description 
\be
X_j= \tfrac{1}{d}\big(\mbb{1} + C_d \bv{x}_j \cdot \bs{\Lambda}\big),
\ee
with $\sum_j\bv{x}_j={\bv 0}$ and $\bv{x}_i \cdot \bv{x}_j = (\delta_{ij}\,d - 1)/(d-1)$, which follow from $\sum_jX_j=\mbb{1}$ and $\Tr(X_iX_j)=\delta_{ij}$. From the algebra induced by $\Lambda_i$ and pertinent vector products, one may prove that $\Tr[(\bv{r}_1\cdot\bs{\Lambda})(\bv{r}_2\cdot\bs{\Lambda})]=2(\bv{r}_1\cdot\bv{r}_2)$, with $\bv{r}_1,\bv{r}_2\in\mb{R}^{d^2-1}$. This simple formula allows one to show that $||\rho_\bv{r}||^2=\frac{1}{d}[1+(d-1)r^2]$, with $r^2=\bv{r}\cdot\bv{r}$, and $p_{x_i}=\Tr\left(X_i\rho\right)=\frac{1}{d}[1+(d-1)\,\bv{x}_i\cdot\bv{r}]$. In addition, one shows that $r_i=\frac{d}{2C_d}\Tr(\Lambda_i\rho)$. With this formalism, we find
\begin{subequations}
\begin{align}
		& \Phi_X(\rho_\bv{r})=\tfrac{1}{d}(\mbb{1} + C_d \bv{u} \cdot \bs{\Lambda}), & \,\,\, & \bv{u}=\tfrac{d-1}{d}\text{\small $\sum_{j=1}^d$} (\bv{x}_j\cdot\bv{r})\,\bv{x}_j,  \\
		& \Phi_{YX}(\rho_\bv{r})=\tfrac{1}{d}(\mbb{1} + C_d \mathbf{v} \cdot \bs{\Lambda}), & \,\,\, & \bv{v}=\tfrac{d-1}{d}\text{\small $\sum_{k=1}^d$}(\bv{y}_k\cdot\bv{u})\,\bv{y}_k,
\end{align}
\label{Phi_uv}
\end{subequations}
where $\bv{u},\bv{v}\in\mb{R}^{d^2-1}$ and $\bv{y}_i\cdot\bv{y}_j=(\delta_{ij}\,d-1)/(d-1)$. Traceless by hypothesis, the considered observables assume the form $X=\bv{x}\cdot\bs{\Lambda}$ and $Y=\bv{y}\cdot\bs{\Lambda}$, where $\bv{x}=(C_d/d)\sum_jx_j\bv{x}_j$ and $\bv{y}=(C_d/d)\sum_ky_k\bv{y}_k$. The relations \eqref{rho_r}-\eqref{Phi_uv} allow us to speak of the incompatibility 
\be 
\ms{I}_\mbb{C}=H\left(\text{\small $\frac{1+(d-1)\bv{y}_k\cdot\bv{u}}{d}$}\right)-H\left(\text{\small $\frac{1+(d-1)\bv{x}_j\cdot\bv{r}}{d}$}\right)
\label{Id}
\ee 
of the ``geometrical context'' $\mbb{C}=\{\bv{r},\bv{x},\bv{y}\}$. In connection with the noisy state~\eqref{rhoe}, we have $\mc{N}_\epsilon(\mbb{C})=\{\epsilon\bv{r},\bv{x},\bv{y}\}$, for which we find a particularly insightful result for the proportionality ratio:
\be 
\ms{R}_\mbb{C}=\frac{||\bv{u}-\bv{v}||^2}{||\,\bv{u}\,||^2}=1-\frac{||\,\bv{v}\,||^2}{||\,\bv{u}\,||^2}.
\label{Rd}
\ee 
To compute the measurement incompatibility we set $\rho=X_j$, which implies that $\bv{r}=\bv{x}_j=\bv{u}$ and, hence, $\ms{R}_{\mbb{C}_j}=1-||\,\bv{v}_j\,||^2$, with $\bv{v}_j=\frac{d-1}{d}\text{\small $\sum_{k=1}^d$}(\bv{y}_k\cdot\bv{x}_j)\,\bv{y}_k$. It then follows that
\be 
\ms{M}_{\{X,Y\}}=1-\frac{1}{d}\sum_{j=1}^d||\,\bv{v}_j\,||^2=1-\frac{d-1}{d^2}\sum_{j,k=1}^d(\bv{x}_j\cdot\bv{y}_k)^2.
\label{Md}
\ee 
The results \eqref{Id}-\eqref{Md} rephrase incompatibility in terms of the geometry defined by the vectors $\bv{r},\bv{x},\bv{y}$. Here the free contexts \eqref{Cfree} manifest themselves with respect to $\ms{I}_\mbb{C}$ as, for instance, $\mbb{C}_1^\text{free}=\{\bv{r},\bv{x},\bv{y}\}$ with $\bv{x}_j\cdot\bv{y}_k=\frac{d\delta_{jk}-1}{d-1}$ (``parallel operators,'' since one has $\bv{x}\cdot\bv{y}=\frac{1}{2}\sum_ix_iy_i$) and $\mbb{C}_2^\text{free}=\{\bv{y}_k,\bv{x},\bv{y}\}$ with $\bv{x}_j\cdot\bv{y}_l=0$ (``orthogonal operators,'' since $\bv{x}\cdot\bv{y}=0$). Interestingly, as far as $\ms{M}_{X,Y}$ is concerned, we see that it vanishes for parallel (commuting) operators and reaches its maximum for orthogonal (MUB forming) operators, this being the heart of our geometrical interpretation.

The scenario becomes rather simple for generic qubit contexts. By setting $d=2$, $C_d=1$, $\bs{\Lambda}=(\sigma_1,\sigma_2,\sigma_3)$, where $\sigma_{1,2,3}$ are the Pauli matrices, $\bv{x}_j=x_j\bv{x}$, and $\bv{y}_k=y_k\bv{y}$ in the precedent formulas we readily obtain
\begin{subequations}
\begin{align}
&\ms{I}_\mbb{C}=h\left(\text{\small $\frac{1 + (\bv{x}\cdot\bv{y})(\bv{x}\cdot\mathbf{r})}{2}$} \right) - h\left(\text{\small $\frac{1 + (\bv{x} \cdot \mathbf{r})}{2}$}\right), \\
&\ms{M}_{\{X,Y\}}=\ms{R}_\mbb{C}=\ms{R}_{\mbb{C}_j}=1-(\bv{x}\cdot\bv{y})^2,
\label{d=2results}
\end{align}
\end{subequations}
where $h(\nu)=-\nu\ln{\nu} - (1-\nu)\ln{(1-\nu)}$ is the binary Shannon entropy and $\bv{r},\bv{x},\bv{y}\in\mb{R}^3$. Some remarks are in order. First, $\ms{I}_\mbb{C}$ is the only quantity that depends on the state $\rho$ (via $\mathbf{r}$), this being the key aspect characterizing it as a context incompatibility. In particular, this ensures that $\ms{I}_\mbb{C}\to 0$ as $|\mathbf{r}|\to 0$ (decoherence-induced classical limit). The ``large mass'' classical limit, on the other hand, effectively comes via $\bv{x}\cdot\bv{y}\cong 1$, which implements the nondisturbance scenario and implies, via $\Phi_{YX}(\rho)\cong\Phi_X(\rho)$, that $\ms{I}_\mbb{C}\cong 0$ (see the Supplemental Material~\cite{SuppMat} for details). This regime is, of course, equivalent to the free context $\mbb{C}_1^\text{free}$, where $\bv{x}=\bv{y}$ (implying parallel operators, that is, $[X,Y]=0$). The context incompatibility vanishes also when $\bv{x}\cdot\bv{r}=0$---case in which the first measurement $X$ is incompatible with the input state---and monotonically increases with the quantifiers given by Eq.~\eqref{d=2results}. Second, by taking $\tfrac{1}{4} ||\,[X,Y]\,||^2=|\bv{x}\times\bv{y}|^2$ as an estimate for the notion of noncommutativity, we see that the ratios $\ms{R}_\mbb{C}$ and $\ms{R}_{\mbb{C}_j}$, with $\mbb{C}_j=\{X_j,X,Y\}$, and the measurement incompatibility $\ms{M}_{\{X,Y\}}$ are all indistinguishable concepts for qubit contexts. To a certain extent, this can be related to the bidirectional implication reported in Ref.~\cite{Heino2010} between the notions of nondisturbance and commutativity.

{\it Conclusion}. In this paper, we have derived a notion of incompatibility from the fact that no information can be extracted from a premeasured state $\Phi_X(\rho)$ if a compatible observable $Y$ is measured in sequence. A distinctive feature of our approach is that it makes reference to a physical context, $\mbb{C}=\{\rho,X,Y\}$, composed not only of observables (measurements) but also of quantum states. Besides allowing one to describe the disappearance of incompatibility in classical regime, our results associate incompatibility with an information-based task in space-time, rather than an algebraic construction in the Hilbert space. The proposed measure of context incompatibility is easily computable and yet admits a norm-based estimate. Remarkably, the context incompatibility is shown to be a resource, with particular application to a protocol devised to test information leakage, makes contact with the notion of measurement incompatibility, and admits a geometrical interpretation in a vector space of arbitrary dimension. Our results give rise to some noteworthy research lines. The first one concerns the extension of our approach to contexts involving more than two (eventually continuous) observables. The second refers to the use of our easily computable quantifier of measurement incompatibility for MUB searching, a longlasting intricate problem in quantum physics.

{\it Acknowledgments}. The authors acknowledge the Brazilian funding agency CNPq, under the grants 147312/2018-3 (E.M.), 160986/2017-6 (M.F.S.), and 303111/2017-8 (R.M.A.), and the National Institute for Science and Technology of Quantum Information (CNPq, INCT-IQ 465469/2014-0).



\newpage
\onecolumngrid

\centerline{\bf \large Supplemental Material}
\appendix
\section{Identification of free contexts with respect to context incompatibility}

\begin{propn}
With respect to contexts $\mbb{C}=\{\rho,X,Y\}$, $\ms{I}_\mbb{C}=0$ if and only if (i) $[X,Y]=0$ $(\forall\rho)$ or (ii) $I^i=I(\Phi_X(\rho))=0$.
\end{propn}

\proof From $\ms{I}_\mbb{C}=S(\Phi_X(\rho)||\Phi_{YX}(\rho))$ it follows that $\ms{I}_\mbb{C}=0$ if and only if $\Phi_{YX}(\rho)=\Phi_X(\rho)$. By its turn, this condition requires, via the definitions of $\Phi_X$ and $\Phi_{YX}$, that 
\be
\sum_ip_j\left[\Phi_Y(X_j)-X_j \right]=0, 
\label{YX=X}
\ee
with $p_j=\Tr(X_j\rho)\geqslant 0$. There are only two ways of satisfying the above equation. First, if $\Phi_Y(X_j)=X_j$ for all $\rho$. This is equivalent to $\sum_kY_kX_jY_k=X_j$.  Multiplying by $Y_l$ either on the left-hand side or on the right-hand side, we obtain $Y_lX_jY_l=Y_lX_j$ or $Y_lX_jY_l=X_jY_l$, respectively, which imply $[X_j,Y_l]=0$ and, hence,  $[X,Y]=\sum_{j,k}x_jy_k[X_j,Y_k]=0$. On the other hand, if $[X,Y]=0$, then $X$ and $Y$ share the same set of projectors, that is, $X_i=Y_i$, which implies that $Y_kX_j=X_jY_k$. Multiplying by $Y_k$ on the right-hand side and summing over $k$, we get $\sum_kY_kX_jY_k=X_j$, which satisfies Eq.~\eqref{YX=X} and completes the proof of item (i). The second way of satisfying Eq.~\eqref{YX=X} is by picking a uniform distribution, $p_j=1/d$, for in this case we find $\tfrac{1}{d}\sum_j[\Phi_Y(X_j)-X_j]=\tfrac{1}{d}[\Phi_Y(\mbb{1})-\mbb{1}]=0$. Since $\Phi_X(\rho)=\sum_jp_jX_j$, in this case we have $\Phi_X(\rho)=\mbb{1}/d$ and $I^i=I(\Phi_X(\rho))=\ln{d}-S(\Phi_X(\rho))=0$. Notice that $\Phi_X(\rho)=\mbb{1}/d$ if, for instance, $\rho=\mbb{1}/d$ $(\forall X,Y)$ or $\rho=\epsilon Y_k+(1-\epsilon)\mbb{1}/d$ (with the $X$ and $Y$ eigenbases forming MUB and $\epsilon\in[0,1]$). Conversely, $I^i=0$ only if $\Phi_X(\rho)=\mathbbm{1}/d$, implying that $\Phi_{YX}(\rho)=\Phi_X(\rho)$ and, hence, the validity of Eq.~\eqref{YX=X}. This proofs item (ii). \hfill $\blacksquare$

\section{Generalized Bloch sphere representation}

Here we briefly review the main aspects of the generalized Bloch sphere representation (see Refs.~\cite{Aerts2014,Aerts2016} for a detailed presentation). Consider a $d$-dimensional Hilbert space $\mc{H}\simeq\mb{C}^d$ and let $\Lambda_j$ be a $d \times d$ complex, traceless, Hermitian, and normalized matrix such that $\Tr(\Lambda_i \Lambda_j)=2 \delta_{ij}$. The matrices $\Lambda_j$ are called the generators of SU($d$), the special group of degree $d$, and, along with the identity matrix, constitute the set $\{\mbb{1},\Lambda_1,\ldots,\Lambda_{d^2-1}\}$, which form an orthogonal basis for all linear operators acting on $\mc{H}$. Since the commutators $[\Lambda_i,\Lambda_j]$ and anticommutators $[\Lambda_i,\Lambda_j]_+$ are self-adjoint operators, they can be expanded in terms of the generators, that is,
\be 
[\Lambda_i,\Lambda_j]=2i\sum_{k=1}^{d^2-1}f_{ijk}\Lambda_k,\qquad\qquad [\Lambda_i,\Lambda_j]_+=\frac{4}{d}\delta_{ij}\mbb{1}+2\sum_{k=1}^{d^2-1}d_{ijk}\Lambda_k, 
\ee 
where the structure constants $f_{ijk}=\tfrac{1}{4i}\Tr\big([\Lambda_i,\Lambda_j]\Lambda_k\big)$ and $d_{ijk}=\tfrac{1}{4}\Tr\big([\Lambda_i,\Lambda_j]_+\Lambda_k\big)$ are elements of a totally antisymmetric tensor and a totally symmetric tensor, respectively. For the two-dimensional case $(d=2)$, the $2^2-1=3$ generators $\Lambda_i$ are the Pauli matrices $\sigma_{1,2,3}$, whereas for $d=3$ the 8 generators are the Gell-Man matrices $\lambda_i$~\cite{Aerts2014,Aerts2016}. A generic state in this representation can be parametrized as
\be
\rho_\bv{r}=\frac{1}{d}\Big(\mbb{1}+C_d \bv{r}\cdot\bs{\Lambda}\Big),
\label{genericstate}
\ee
where $\mathbf{r}=\sum_{i=1}^{d^2-1}r_i\bv{e}_i$, $C_d \equiv \sqrt{d(d-1)/2}$, $\bs{\Lambda} = \sum_{i=1}^{d^2-1}\Lambda_i\bv{e}_i$, and $\{\bv{e}_i\}_{i=1}^{d^2-1}$ is an orthonormal basis of $\mb{R}^{d^2-1}$. By introducing the star and wedge products, which are respectively defined as 
\be 
(\bv{u}\star\bv{v})_i=\frac{C_d}{d-2}\sum_{j,k=1}^{d^2-1}d_{ijk}u_jv_k, \qquad\qquad (\bv{u}\wedge\bv{v})_i=\sum_{j,k=1}^{d^2-1}f_{ijk}u_jv_k, 
\ee 
with $\bv{u}\star\bv{v}=\bv{v}\star\bv{u}$ and $\bv{u}\wedge\bv{v}=-\bv{v}\wedge\bv{u}$, one can show that
\be 
(\bv{u}\cdot\bs{\Lambda})(\bv{v}\cdot\bs{\Lambda})=\frac{2}{d}(\bv{u}\cdot\bv{v})\mbb{1}+\imath\,(\bv{u}\wedge\bv{v})\cdot\bs{\Lambda}+\frac{d-2}{C_d}(\bv{u}\star\bv{v})\cdot\bs{\Lambda}.
\ee 
This allows us to deduce the important relation $\Tr\left[(\bv{u}\cdot\bs{\Lambda})(\bv{v}\cdot\bs{\Lambda})\right]=2(\bv{u}\cdot\bv{v})$. With these formulas, one can compute the norm
\be 
||\,\rho_\bv{r}\,||^2=\Tr\rho_\bv{r}^2=\frac{1}{d}\Big[1+(d-1)r^2\Big],
\label{norm}
\ee
where $r=||\mathbf{r}||=\sqrt{\bv{r}\cdot\bv{r}}$, meaning that pure (mixed) states have $r=1$ ($r<1$).
We can obtain the components of $\bv{r}$ from the density operator and the set of generators as $r_i=\frac{d}{2C_d}\Tr(\rho\Lambda_i)$, which shows that the state can be reconstructed from the expectation values of the generators, thus making $\{\Lambda_1,\ldots,\Lambda_{d^2-1}\}$ a set of informationally complete observables~\cite{Prugovecki1977, Carmeli2012}.

Projection operators assume the form $X_i = \frac{1}{d}(\mbb{1} + C_d \bv{x}_i \cdot \bs{\Lambda})$, with $\sum_i\bv{x}_i=0$ and $\bv{x}_i \cdot \bv{x}_j = (\delta_{ij}\,d - 1)/(d-1)$, which follow from $\sum_iX_i=\mbb{1}$ and $\Tr(X_iX_j)=\delta_{ij}$ [plus the relation \eqref{norm}], respectively. If $X$ is traceless, then $X=\sum_ix_iX_i=\bv{x}\cdot\bs{\Lambda}$, where $\bv{x}=(C_d/d)\sum_ix_i\bv{x}_i$ and $\bv{x}\cdot\bv{x}=\tfrac{1}{2}\sum_ix_i^2$, with $x_i$ being the eigenvalues of $X$. Probability distributions are computed as
\be
p_i=\Tr\left(X_i\rho_\bv{r}\right)=\frac{1}{d}\Big[1+(d-1)\,\bv{x}_i\cdot\bv{r}\Big].
\label{p_xi}
\ee

Using these tools, we can calculate the post-measurement state $\Phi_X(\rho_\bv{r})$, for an observable $X = \sum_j x_j X_j$. Notice that $\Phi_X(\rho_\bv{r})$ must be of the same form as Eq.~\eqref{genericstate}, with the only difference coming from the vector that represents it. Indeed, we arrive at
\be
\Phi_X(\rho_\bv{r})=\sum_{j=1}^d \Tr[X_j\rho_\bv{r}] X_j = \frac{1}{d}\Big[\mbb{1} + C_d \bv{u} \cdot \bs{\Lambda}\Big],\qquad\qquad \bv{u}\coloneqq\frac{d-1}{d} \sum_{j=1}^d (\bv{x}_j\cdot\bv{r})\,\bv{x}_j,
\label{PhiX}
\ee
where $\bv{u}$ is the vector in $\mb{R}^{d^2-1}$ representing the resulting state. We see that this transformation selects only the projections of {\bf r} into the $\bv{x}_j$ axes, thus rotating and also contracting it, since $||\bv{u}|| \leq r$, with equality applying when $\rho_\bv{r}$ is an eigenvector of $X$. In fact, for $\rho_\bv{r}=X_i$, we have $\bv{r}=\bv{x}_i$, which implies $\bv{u}=\bv{x}_i$, and $\Phi_X(\rho_\bv{r})=X_i$. If we now perform the second measurement, the same arguments hold and we get
\be
\Phi_{YX}(\rho_\bv{r}) = \sum_{k=1}^d \Tr\left[Y_k\Phi_X(\rho_\bv{r})\right] Y_k = \frac{1}{d}\Big[\mbb{1} + C_d \mathbf{v} \cdot \bs{\Lambda}\Big],\qquad\qquad \bv{v}\coloneqq\frac{d-1}{d}\sum_{k=1}^d(\bv{y}_k\cdot\bv{u})\,\bv{y}_k,
\label{PhiYX}
\ee
where $\bv{y}_i\cdot\bv{y}_j=(\delta_{ij}\,d-1)/(d-1)$. Since the eigenvalues of $\Phi_X(\rho_\bv{r})$ and $\Phi_{YX}(\rho_\bv{r})$ are given by Eq.~\eqref{p_xi} and $\frac{1}{d}\big[1+(d-1)\bv{y}_k\cdot\bv{u}\big]$, respectively, we can compute the context incompatibility [Eqs.(4) and (14) of the main text]:
\be 
\ms{I}_\mbb{C}=H\left(\frac{1+(d-1)\bv{y}_k\cdot\bv{u}}{d}\right)-H\left(\frac{1+(d-1)\bv{x}_j\cdot\bv{r}}{d}\right),
\ee 
where $H(\nu_i)=-\sum_i\nu_i\ln{\nu_i}$ is the Shannon entropy of the distribution $\nu_i$. The results \eqref{PhiX} and \eqref{PhiYX} allow us to obtain an insightful result for the proportionality ratio [see Eqs. (7) and (15) of the main text]:
\be 
\ms{R}_\mbb{C}=\frac{||\Phi_X(\rho)-\Phi_{YX}(\rho)||^2}{||\Phi_X(\rho)-\mbb{1}/d||^2}=\frac{||\Phi_X(\rho)||^2-||\Phi_{YX}(\rho)||^2}{||\Phi_X(\rho)||^2-1/d}=\frac{||\bv{u}-\bv{v}||^2}{||\,\bv{u}\,||^2}=1-\frac{||\,\bv{v}\,||^2}{||\,\bv{u}\,||^2}.
\ee 
In particular, for the context $\mbb{C}_j=\{X_j,X,Y\}$, for which $\bv{u}=\bv{x}_j$, the above formula reduces to
\be 
\ms{R}_{\mbb{C}_j}=1-||\,\bv{v}_j\,||^2=1-\frac{d-1}{d}\sum_{k=1}^d(\bv{x}_j\cdot\bv{y}_k)^2=\frac{d}{d-1}\left(1-\sum_{k=1}^d|\braket{x_j|y_k}|^4\right).
\label{Rjd}
\ee 
Then, we can write our measurement incompatibility quantifier [see Eqs.(9) and (16) of the main text] as
\be 
\ms{M}_{\{X,Y\}}=1-\frac{1}{d}\sum_{j=1}^d||\,\bv{v}_j\,||^2=1-\frac{d-1}{d^2}\sum_{j,k=1}^d(\bv{x}_j\cdot\bv{y}_k)^2=\frac{1}{d-1}\left(d-\sum_{j,k=1}^d|\braket{x_j|y_k}|^4\right).
\label{MXYd}
\ee 
This result is identical to the measure employed as a quantifier of mutual unbiasedness~\cite{Bengtsson2007,Durt2010}. This idea is supported by the fact that $0\leq \ms{M}_{X,Y}\leq 1$, where the upper (lower) bound is reached if and only if $|\braket{x_j|y_k}|^2=1/d$ ($|\braket{x_j|y_k}|^2=\delta_{x_jy_k}$), which amounts to having $X$ and $Y$ as MUB (commuting operators). In the present work, however, this is viewed as a demonstration that $\ms{M}_{\{X,Y\}}$ is an incompatibility measure related solely to the observables $X$ and $Y$, being, in this capacity, a quantifier of measurement incompatibility. In the next section, we show that $\ms{M}_{\{X,Y\}}$ can be constructed via an purely algebraically-oriented way.

For qubits $(d=2)$, the above results assume the simple forms presented in the main text.

\section{Purely algebraic construction of the measurement incompatibility quantifier $\ms{M}_{\{X,Y\}}$}

This section aims at constructing a measure intended to capture the interconnections between the algebraic properties of the observables $X$ and $Y$. Accordingly, the resulting measure is expected to be sensitive to the commutation relation of these observables and to the structure of their eigenbases. Let us start by introducing the matrices
\be 
\Pi_X \stackrel{\cdot}{=} \left[\begin{matrix} X_1 \\ X_2 \\ \vdots \\ X_d \end{matrix}\right],\qquad\qquad 
\Pi_Y \stackrel{\cdot}{=} \left[\begin{matrix} Y_1 \\ Y_2 \\ \vdots \\ Y_d \end{matrix}\right],\qquad\qquad
\mbb{M}_{XY}:=\Pi_X\Pi_Y^T \stackrel{\cdot}{=} \left[\begin{matrix} X_1Y_1 & X_1Y_2 & \cdots & X_1Y_d \\ X_2Y_1 & X_2Y_2 & \cdots & X_2Y_d \\ \vdots & \vdots & \ddots & \vdots \\ X_dY_1 & X_dY_2 & \cdots & X_dY_d\end{matrix}\right],
\label{PiXY}
\ee 
which are formed with the projectors $X_j=\ket{x_j}\bra{x_j}$ and $Y_k=\ket{y_k}\bra{y_k}$ of $X$ and $Y$, respectively. The $d\times d$ matrix $\mbb{M}_{XY}$ contains detailed structural information about the underlying algebra of the pair $\{X,Y\}$. Now, one may check that $\mbb{M}_{QQ}=\Pi_Q\Pi_Q^T=\mbb{I}\Pi_Q$ ($Q\in\{X,Y\}$), with $\mbb{I}$ the $d\times d$ identity matrix. It follows that $\mbb{M}_{QQ}\mbb{M}_{QQ}^T=\mbb{M}_{QQ}$ and $\text{tr}(\mbb{M}_{QQ})=\sum_iQ_i=\mbb{1}$, from which we have $\Tr[\text{tr}(\mbb{M}_{QQ})]=d$ (we use ``tr'' to denote the trace operation over the matrix algebra presently introduced). On the other hand, $\mbb{M}_{XX}\mbb{M}_{YY}^T=\mbb{M}_{XY}$. We then compute the extent to which the matrix $\mbb{M}_{XY}$ is different from $\mbb{M}_{QQ}$ through the measure
\beq 
\ms{D}_{\{X,Y\}}&=&||\mbb{M}_{QQ}||^2-||\mbb{M}_{XY}||^2=d-\Tr\left\{\text{tr}\left[\mbb{M}_{XY}\mbb{M}_{XY}^T \right]\right\}\nonumber \\
&=& d-\Tr\sum_{j,k=1}^dX_jY_kX_jY_k=d-\sum_{j,k=1}^d|\braket{x_j|y_k}|^4.
\eeq
In connection with the result~\eqref{MXYd}, one finds
\be
\ms{M}_{\{X,Y\}}=\frac{\ms{D}_{\{X,Y\}}}{d-1}.
\ee
This result demonstrates how the measure $\ms{M}_{\{X,Y\}}$, derived in the main text through a protocol involving information leakage, is in full consonance with an algebraically-oriented formulation that is devised to encode the structural relations between the observables eigenbases. Clearly, $\ms{D}_{\{X,Y\}}$ trivially inherits all the proprerties pointed out in the main text for $\ms{M}_{\{X,Y\}}$.

\section{The ``large-mass limit'' of context incompatibility}

Suppose that, after being prepared in a generic state [see Eq.~(11) with $d=2$], an electron is submitted to a sequence of two Stern-Gerlach (SG) magnets, one with magnetic field $B_x\bv{x}$ and the other with $B_y\bv{y}$.  Here the scenario is such that the physical interaction generates spin-position correlations that perfectly induce measurements of the observables $X=\bv{x}\cdot\bs{\sigma}$ and $Y=\bv{y}\cdot\bs{\sigma}$. That is, the electron spin couples with the magnetic fields in a way such that one can clearly distinguish between displacements of the electron trajectory along the axes $x$ or $y$. In this case, the $X$ and $Y$ bases form MUB and $\bv{x}\cdot\bv{y}=0$, which maximizes both the context incompatibility and the resulting measurement incompatibility [see Eqs.~(17)]. Now, consider that the same SG apparatuses are used to measure the spin of an extremely massive particle. In this case, the coupling with the magnetic field is not enough to cause a significant deviation in the particle trajectory. In other words, the correlations established between position and spin are tiny, and the resulting ``measurement'' becomes effectively fuzzy. The larger the mass, the greater the difficulty to experimentally distinguish between the observables $\bv{x}$ and $\bv{y}$. As a consequence, $\bv{x}\cong \bv{y}$ and $\ms{I}_\mbb{C}\cong 0$.

But our approach admits an even deeper treatment of this problem. To make the point, let us imagine that the SG magnets are free to move upon interaction with the passing particle, that is, the apparatuses can receive kickbacks that guarantee total momentum conversation. Being free to move, the magnets earn the right to be treated quantum mechanically. In this case, via the Stinespring theorem, the unrevealed-measurement map is written $\Phi_X(\rho)=\Tr_\text{\tiny SG}\big(U\rho\otimes\rho_\text{\tiny SG} U^\dag \big)$, with $\rho_\text{\tiny SG}$ being the apparatus state. At this stage, the notion of collapsing measurement associated with the projectors $X_j$ is replaced with correlations created between the systems by $U$. When the mass of the particle is large and the coupling with the field is weak, the correlations are nearly negligible, which legitimates us to replace the projective-measurement map $\Phi_X$ with the weak-measurement map~\cite{Dieguez2018} 
\be 
M_X^\epsilon(\rho)\coloneqq (1-\epsilon)\,\rho+\epsilon\,\Phi_X(\rho),
\ee
where $0\leq\epsilon\leq 1$. Being related to the effective strength of the measurement, $\epsilon$ can be shown~\cite{Dieguez2018} to emerge from the  strength of the physical coupling. In the problem under scrutiny here, since the interaction is momentum conserving, then we can expect that $\epsilon\propto m_\text{\tiny SG}/m$, where $m$ $(m_\text{\tiny SG})$ is the mass of the particle (SG magnet). Upon the replacement $\Phi\to M^\epsilon$, context incompatibility becomes $\ms{I}_\mbb{C}=S\big(M_Y^\epsilon M_X^\epsilon(\rho)\big)-S\big(M_X^\epsilon(\rho)\big)$. In the regime where $\epsilon\ll 1$, we have
\be 
M_Y^\epsilon M_X^\epsilon(\rho)\cong M_X^\epsilon(\rho)+\epsilon\big[\Phi_Y(\rho)-\rho \big],
\ee 
which allows us to conclude that the large-mass limit, $(m_\text{\tiny SG}/m)\to 0$, implies that $\ms{I}_\mbb{C}\to 0$. When the particle is lightweight in comparison with the magnets, then the model can be improved as $\epsilon\propto 1-\exp\left(-m_\text{\tiny SG}/m \right)$, which correctly yields $\epsilon\to 1$ for $(m_\text{\tiny SG}/m)\to\infty$ and also encapsulates the large-mass limit. Although the discussion has been conducted here for $d=2$, similar arguments apply in general.



\end{document}